\documentclass[12pt]{article}
\usepackage{epsf}
\usepackage{amsmath}
\usepackage{graphics}
\usepackage{cite}
\usepackage{color}
\input{colordvi.tex}

\renewcommand{\thefootnote}{\#\arabic{footnote}}

\begin{document}

\setcounter{footnote}{0}
\begin{titlepage}

\begin{center}

\hfill May 2008\\

\vskip .5in

{\Large \bf
Primordial Helium Abundance from CMB: \\
a constraint from recent observations and a forecast}

\vskip .45in

{\large
Kazuhide Ichikawa$^{1,2}$,
Toyokazu Sekiguchi$^1$,
and Tomo Takahashi$^3$ 
}

\vskip .45in

{\em
$^1$
Institute for Cosmic Ray Research, 
University of Tokyo, Kashiwa 277-8582, Japan\\
$^2$
Department of Physics and Astronomy, University College London, Gower Street, London, WC1E 6BT, U.K. \\
$^3$
Department of Physics, Saga University, Saga 840-8502, Japan  

}

\end{center}

\vskip .4in

\begin{abstract}
  We studied a constraint on the primordial helium abundance $Y_p$ from
  current and future observations of CMB.  Using the currently
  available data from WMAP, ACBAR, CBI and BOOMERANG, we obtained the
  constraint as $Y_p = 0.25^{+0.10}_{-0.07}$ at 68\% C.L.  We also
  provide a forecast for the Planck experiment using the Markov chain Monte Carlo
  approach.  In addition to forecasting the constraint on $Y_p$, we
  investigate how assumptions for $Y_p$ affect constraints on the
  other cosmological parameters.
 
\end{abstract}
\end{titlepage}

\renewcommand{\thepage}{\arabic{page}}
\setcounter{page}{1}
\renewcommand{\thefootnote}{\#\arabic{footnote}}

\section{Introduction}

Current cosmological observations are very precise to give us a lot of
information on the evolution and present state of the universe.  We
usually extract the information by constraining cosmological
parameters such as energy densities of baryon, dark matter and dark
energy, the Hubble constant, reionization optical depth, spectral
index of primordial fluctuation and so on. Among them, in this paper,
we focus on the primordial helium abundance $Y_p$, which has been of
great interest in cosmology.  One of the reasons why the primordial
helium abundance has been considered to be interesting and important
is that, in the context of the standard big bang nucleosynthesis
(SBBN), we can know the baryon density once $Y_p$ is determined from
observations.  However, it has been discussed that a significant
systematic error dominates when one infers the value of $Y_p$ from
measurements in low-metallicity extragalactic HII region
\cite{Olive:2004kq,Cyburt:2004yc,Fukugita:2006xy,
  Peimbert:2007vm,Izotov:2007ed}.  Furthermore, there have been some
discussions that there may be a large uncertainty in the neutron
lifetime \cite{Serebrov:2004zf,Mathews:2004kc,Serebrov:2006im}, which
results in uncertainties in the predictions for the abundances of
light elements.  In this respect, the study of other independent
measurements of the helium abundance would be interesting.

Recent precise measurements of cosmic microwave background (CMB) such
as WMAP can now enable us to constrain cosmological parameters with
great accuracies.  However, the helium abundance has not been
discussed much in the literature when one study of cosmological
constraints from CMB since $Y_p$ has been considered to have little
effect on CMB power spectrum. In most of analyses, $Y_p$ is fixed to
be $0.24$ which is probably motivated from a somewhat old value of the
observed primordial helium abundance of $Y_p= 0.238$
\cite{Fields:1998gv}\footnote{
  Recent observations give e.g. $Y_p = 0.249 \pm 0.009$
  \cite{Olive:2004kq,Cyburt:2004yc}.
}.  But, in fact, there have been some works which discuss the effects
of $Y_p$ on CMB and some constraints were given
\cite{Trotta:2003xg,Huey:2003ef,Ichikawa:2006dt}.  Since the helium
abundance affects the recombination history, the CMB power spectrum
can be affected mainly through the diffusion damping.  Although the
constraint is not so severe, it is important to notice that they are
obtained independently from BBN, which can be used to cross-check our
understanding of the helium abundance.  Furthermore, the value of
$Y_p$ at the time of BBN may be different from that at late time when
CMB observations can probe\footnote{
  For example, in a scenario to solve so-called ``Lithium problem"
  with $Q$-balls, the decay of $Q$-balls produce extra baryon after
  BBN has completed \cite{Ichikawa:2004pb}.  In a model of this kind,
  the value of $Y_p$ can vary at different epochs.
}. After we studied the constraint on $Y_p$ in \cite{Ichikawa:2006dt},
the data from WMAP has been updated
\cite{Komatsu:2008hk,Dunkley:2008ie,Hinshaw:2008kr,Hill:2008hx,Nolta:2008ih}.
In addition, the data at higher multipoles where the effects of $Y_p$
become significant have been updated by ACBAR \cite{Reichardt:2008ay}
and CBI \cite{Sievers:2005gj}. Thus it is a good time to investigate the
constraint on the helium abundance using these CMB data, which is one
of the aims of this paper.

Furthermore, we expect a more precise measurement of CMB from the
future Planck satellite \cite{:2006uk}, which can give us a much
better constraint on $Y_p$. In fact, future constraint on $Y_p$ has
already been studied using the Fisher matrix formalism
\cite{Trotta:2003xg,Ichikawa:2006dt}. Although this method is fast and
usually adopted to predict the precision of the future measurements of
cosmological parameters, it can give some inaccurate predictions in
some cases, for example, when the likelihoods does not respect a
Gaussian form.  In addition, the Fisher matrix formalism predicts only
the uncertainty for the parameter estimation since it just concerns
with the derivatives with respect to parameters around the fiducial
values.  However, since some parameters are correlated in general,
fixing the values of some parameters can bias the estimation of other
parameters.  Namely, priors we assume on some parameters can cause the
estimated central values to deviate from the input fiducial values,
but such effects cannot be quantified by the Fisher matrix approach.
Thus, in this paper, we use the Markov chain Monte Carlo (MCMC) approach
to extract reliable future constraints on $Y_p$ and other
cosmological parameters. In particular, when we forecast the
sensitivity for other cosmological parameters, we assume some
different priors on $Y_p$ and investigate to what extent the
information of $Y_p$ is important to determine other cosmological
parameters.

The organization of this paper is as follows. In the next section, the
effects of $Y_p$ on CMB power spectrum are briefly discussed.  Then in
section \ref{sec:current}, we study the constraint on $Y_p$ using CMB
data currently available including ACBAR, BOOMERANG and CBI as well as
WMAP5.  In section \ref{sec:Planck}, we investigate a constraint on
$Y_p$ and other cosmological parameters from the future Planck experiment.
In addition, we also study how the prior on $Y_p$ affects the
determination of other cosmological parameters.  A brief
  discussion on the significance of the uncertainties of the recombination
  theory in deriving cosmological constraint is given too.   The
final section is devoted to the summary of this paper.

%%%%%%%%%%%%%%%%%%%%%%%%%%%%%%
\section{Effects of $Y_p$ on CMB}\label{sec:effects}
%%%%%%%%%%%%%%%%%%%%%%%%%%%%%%

\begin{figure}[htb]
\begin{center}
\scalebox{0.8}{\hspace{-0.3cm}
\includegraphics{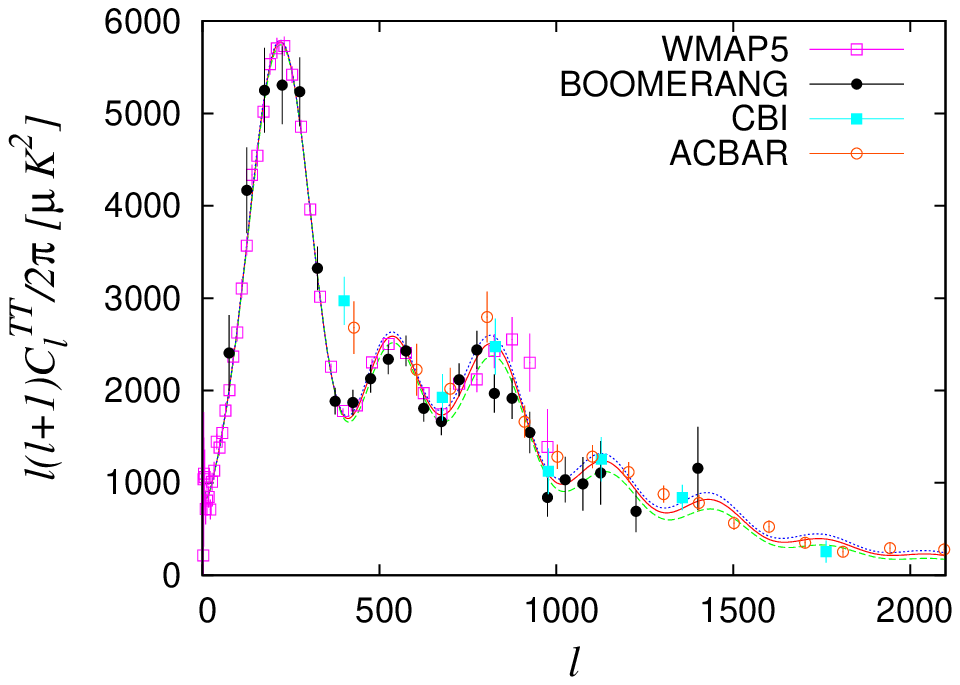}\hspace{-0.5cm}
\includegraphics{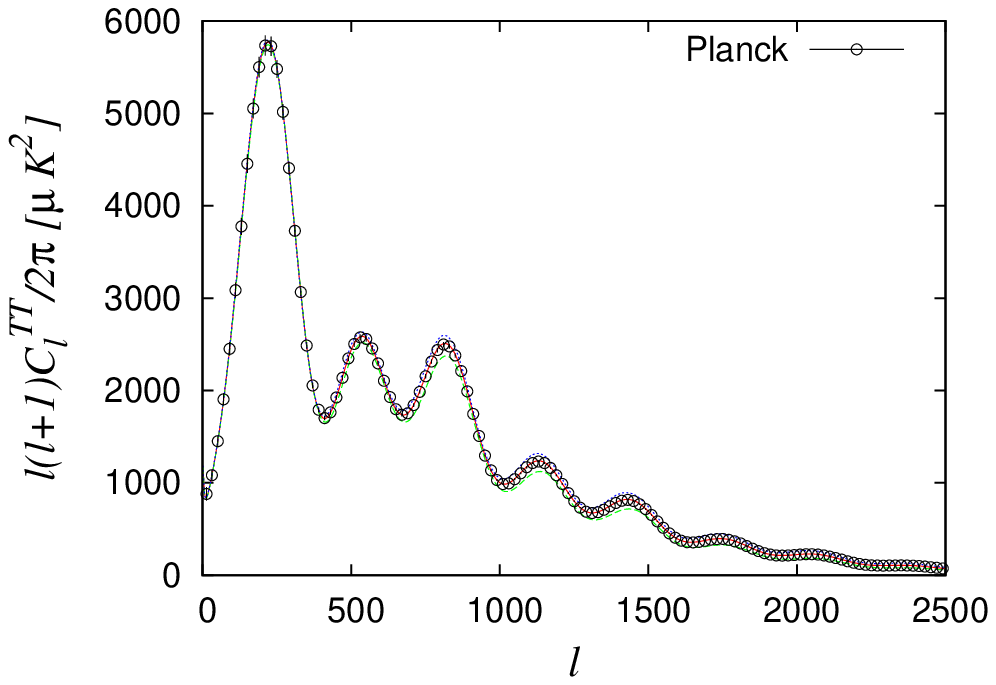}
}
\caption{CMB $TT$ power spectra for several values of $Y_p$.  In this
  figure, we take $Y_p = 0.1$ (blue dotted line), $0.24$ (red solid
  line) and $0.4$ (green dashed line). 
  Other cosmological
  parameters are assumed to be the mean value of WMAP5 for a
  power-law $\Lambda$CDM model. 
  For reference, in the left panel,
  data from WMAP5, ACBAR, BOOMERANG and CBI are also depicted.  In the
  right panel, the expected data from the Planck experiment are also shown. }
\label{fig:cl}
\end{center}
\end{figure}

Here we briefly discuss the effects of the primordial helium abundance $Y_p$
on CMB power spectrum where $Y_p = 4 n_{\rm He}/(n_{\rm H}+ 4 n_{\rm He})$ with $n_{\rm H}$
and $n_{\rm He}$ being the number density of hydrogen and helium-4 
respectively. As has been discussed in
\cite{Trotta:2003xg,Ichikawa:2006dt}, the value of $Y_p$ can affect
the recombination history, which changes the structure of acoustic
peaks.  The effects of $Y_p$ on acoustic peaks mainly come from the
diffusion damping which suppresses the power on small scales and the
shift of the position of acoustic peaks due to the change of the
recombination epoch.  Before recombination, the number density of
electron $n_e$ can be given by $n_e = n_b ( 1 - Y_p)$, where $n_b$ is
the baryon number density.  Thus the increase of $Y_p$ indicates the
decrease of the number of electrons.  When the number of electrons is
reduced, the mean free path of the Compton scattering becomes larger,
which means that fluctuations on larger scales can be more affected by
the diffusive mixing and rescattering.  Thus the damping scale below
which fluctuation of photon is exponentially suppressed becomes
larger.  Furthermore, due to the change of the number density of
electrons, the epoch of recombination is also affected even though its
effect is not so significant. This effect shifts the position of
acoustic peaks slightly.  In Fig.~\ref{fig:cl}, we show the CMB $TT$
spectrum with several values of $Y_p$.  For reference, we also plot
the current data (left panel) and the expected data from the future Planck
experiment (right panel).  As mentioned above, by increasing the value
of $Y_p$, the power on small scales is damped more.  In addition, it
is noticeable that the position of acoustic peaks also shifts
slightly.

To characterize the effects of the change in $Y_p$ and other
cosmological parameters on CMB $TT$ power spectrum $C^{TT}_l$, we
consider some useful quantities \cite{Hu:2000ti}.  First of all, to
see how cosmological parameters affect the position of acoustic peaks,
we investigated the response of the position of the first peak by the
change of the parameters, which we denote $\Delta l_1$.  In addition,
to see the effects of the diffusion damping and some other effects by
cosmological parameters, we study the height of the first peak
relative to that at $l=10$ and the height of the second peak (and
higher peaks up to 5th peak) relative to the first peak, which are
denoted as $H_1$, $H_2$, $H_3$, $H_4$ and $H_5$, respectively.  For
clarity, we give the definitions of these quantities.  The definition
of $H_1$ is
\begin{equation}
H_1 \equiv \left( \frac{\Delta T(l=l_1)}{\Delta T(l=10)} \right)^2,
\end{equation}
and the height of the $i$-th peak relative to the first peak is
defined as
\begin{equation}
H_i \equiv \left( \frac{\Delta T(l=l_i)}{\Delta T(l=l_1)} \right)^2 ~~( {\rm for}~ i \ge 2),
\end{equation}
where $(\Delta T(l))^2 = l(l+1)C^{TT}_l /2\pi$.  We varied
cosmological parameters including $Y_p$ around a fiducial model and
obtained partial derivatives by fitting linearly around the fiducial
value.  For the fiducial cosmological values, we assumed the mean
values of the WMAP 5-yr result (WMAP5) for a power-law $\Lambda$CDM model.  Regarding $Y_p$,
we take $Y_p=0.248$ which corresponds to the value obtained in the
SBBN for the WMAP5 baryon density.  The resulting derivatives are:
\begin{eqnarray}
\Delta l_1 &=&
15.6\frac{\Delta \omega_b}{\omega_b} 
-27.0\frac{\Delta \omega_m}{\omega_m} 
+36.0\frac{\Delta n_s}{n_s} 
+0.94  \frac{\Delta Y_p}{Y_p}
-44.5 \frac{\Delta h}{h},
\label{eq:L1} 
\\
\Delta H_1 &=& 
2.87\frac{\Delta \omega_b}{\omega_b} 
-3.13\frac{\Delta \omega_m}{\omega_m} 
+16.7\frac{\Delta n_s}{n_s} 
-2.30 \frac{\Delta h}{h} , 
 \label{eq:H1}  \\
\Delta H_2 &=& 
-0.290\frac{\Delta \omega_b}{\omega_b} 
+0.023\frac{\Delta \omega_m}{\omega_m} 
+0.396\frac{\Delta n_s}{n_s} 
-0.013 \frac{\Delta Y_p}{Y_p},
\label{eq:H2} \\
\Delta H_3 &=&
-0.177\frac{\Delta \omega_b}{\omega_b} 
+0.206\frac{\Delta \omega_m}{\omega_m} 
+0.514\frac{\Delta n_s}{n_s} 
-0.028 \frac{\Delta Y_p}{Y_p}, 
\label{eq:H3} \\
\Delta H_4 &=& 
-0.102\frac{\Delta \omega_b}{\omega_b} 
+0.082\frac{\Delta \omega_m}{\omega_m} 
+0.317\frac{\Delta n_s}{n_s} 
-0.025 \frac{\Delta Y_p}{Y_p}, 
\label{eq:H4} \\
\Delta H_5 &=& 
-0.040\frac{\Delta \omega_b}{\omega_b} 
+0.084\frac{\Delta \omega_m}{\omega_m} 
+0.236\frac{\Delta n_s}{n_s} 
-0.023 \frac{\Delta Y_p}{Y_p},
\label{eq:H5} 
\end{eqnarray}
where $\omega_b$ and $\omega_m$ are energy densities of baryon and
matter, $n_s$ is the scalar spectral index of primordial fluctuation,
$h$ is the Hubble constant in units of $100~{\rm km}~{\rm s}^{-1}{\rm
  Mpc}^{-1}$.  In the formula for $H_1$, we do not show the dependence
on $\Delta Y_p/ Y_p$ since its effect on $H_1$ is very small compared
to that of the other parameters.  As seen from the negative signs of
$\Delta H_i/\Delta Y_p$ for $i=2$-5, the diffusion damping becomes
more efficient as $Y_p$ increases.  We can also see the correlation of
$Y_p$ with other cosmological parameters, which can be useful when we
interpret the results, in particular, for a Planck forecast.

%%%%%%%%%%%%%%%%%%%%%%%%%%%
\section{Current constraint on $Y_p$}\label{sec:current}
%%%%%%%%%%%%%%%%%%%%%%%%%%%

\begin{table}[htb]
  \begin{center}
  \begin{tabular}{l||c|c}
  \hline
  \hline
  &\multicolumn{2}{c}{prior ranges} \\
  parameters & current data & Planck\\
  \hline
  $\omega_b$ & $0.005\to0.1$& $0.005\to0.1$\\
  $\omega_c$ & $0.01\to0.99$& $0.01\to0.99$\\
  $\theta_s$ & $0.5\to10$& $0.5\to10$\\
  $\tau$ & $0.01\to0.8$& $0.01\to0.8$\\
  $n_s$ & $0.5\to1.5$& $0.5\to1.5$\\
  $\ln(10^{10}A_s)$ & $2.7\to4$& $2.7\to4$\\
  $Y_p$ & $(0\to1)$& $(0\to1)$\\
  $A_\mathrm{SZ}$ & $0\to2$ & ---\\
  $F_H$ & --- & ($0\to2$)\\
  $b_{He}$ & --- & ($0\to1.5$)\\  
  \hline
  \hline 
\end{tabular}
\caption{ The prior ranges for the parameters used in the analysis.
  Priors shown in the 1st and 2nd columns are adopted for current
  and expected Planck data, respectively.  Note that we adopt the
  prior range for $Y_p$ shown above only in the cases with $Y_p$ being
  treated as a free parameter whereas $Y_p$ is a derived parameter in the case where we
  assume the SBBN relation.  For the
  analysis with the current data, we also vary the amplitude of the
  Sunyaev-Zel'dovich effect $A_{\rm SZ}$, which is omitted in the Planck data analysis. 
  We also include two
  additional parameters $F_H$ and $b_{He}$ which represent
  uncertainties in the theory of recombination (see
  section~\ref{sec:Planck} for more details).  }
  \label{table:priors}
\end{center}
\end{table}

Now we discuss the constraint on $Y_p$ from current cosmological
observations.  For this purpose, we make use of the CMB data from
WMAP5
\cite{Komatsu:2008hk,Dunkley:2008ie,Hinshaw:2008kr,Hill:2008hx,Nolta:2008ih},
ACBAR \cite{Reichardt:2008ay}, BOOMERANG
\cite{Jones:2005yb,Piacentini:2005yq,Montroy:2005yx} and CBI
\cite{Sievers:2005gj}.
To investigate the
constraint, we performed a MCMC analysis by using a modified version
of {\tt cosmomc} code \cite{Lewis:2002ah}.  We sampled in an 8
dimensional parameter space with $(\omega_b, \omega_c, \tau, \theta_s,
n_s, A_s, Y_p,A_{SZ})$ where $\omega_c$ is the energy density of dark
matter, $\tau$ is the optical depth of reionization, $\theta_s$ is
acoustic peak scale \cite{Kosowsky:2002zt}, $A_s$ is the amplitude of
primordial curvature fluctuation at the pivot scale $k_0=0.05$
Mpc$^{-1}$ and $A_{SZ}$ is the amplitude of thermal Sunyaev-Zel'dovich (SZ) effect 
which is normalized to the $C^{SZ}_l$ template from
\cite{Komatsu:2002wc}\footnote{
However, the SZ effect may be so large at very high multipoles that this template may not be appropriate to adopt. Hence, we conservatively do not use the ACBAR and CBI data with $ l \ge 2100$.
}.
In this paper, we consider a flat universe and
assume a cosmological constant as dark energy. We also assume no
running for primordial scalar fluctuation and no tensor mode.  When we
report our results in the following, we also use other customarily
used cosmological parameters such as the Hubble constant $H_0 = 100
h~{\rm km}~{\rm s}^{-1}{\rm Mpc}^{-1}$ and energy density of matter
$\Omega_m = (\omega_b + \omega_c)/h^2$.  In performing a MCMC
analysis, we impose top-hat priors on the primary parameters given
above, which are summarized in Table~\ref{table:priors}.

\begin{figure}[tb]
\begin{center}
\scalebox{0.8}{\includegraphics{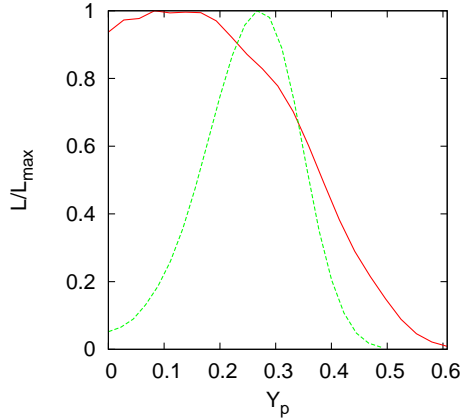}}
\caption{One-dimensional marginalized distribution of $Y_p$ for the
  cases with WMAP5 alone (red solid line) and all CMB data combined
  (green dashed line).  }
\label{fig:Yp}
\end{center}
\end{figure}

Now we discuss the constraints on $Y_p$ when WMAP5 alone is used and
when the data of ACBAR, BOOMERANG and CBI are used in addition.  In
Fig.~\ref{fig:Yp}, we show one-dimensional marginalized distributions
of $Y_p$ for these two cases and, in the 1st and 2nd columns of
Table~\ref{table:current}, we list the parameter estimations for $Y_p$
and other cosmological parameters.  When we use WMAP5 data alone,
the constraint is given as $Y_p \le 0.44$ at 95\%\,C.L. where 
we only give the upper bound since 
the likelihood has a sizable value at $Y_p = 0$.
On the other hand,  when all data are combined,
the constraint is $Y_p = 0.25^{+0.10 (+ 0.15)}_{-0.07 (-0.17)}$
at 68\% (95\%) C.L.
For the analysis using WMAP5 alone,
the limit we obtained is consistent with the result given in
\cite{Dunkley:2008ie}.  
We see that by including the data from ACBAR, BOOMERANG and CBI
the likelihood distribution has a well-defined peak which is close to Gaussian.
 It may also be
interesting to notice that when the data from ACBAR and BOOMERANG and
CBI are combined, the mean value becomes as $Y_p = 0.25$, which is
very close to the value obtained from HII region observations although
the uncertainty is still large.  With the help of this data set, we
may begin to see the concordance with regards the $Y_p$ measurement from
CMB and that from the HII regions.

\begin{table}[ht]
  \begin{center}
  \begin{tabular}{l||r|r|r}
  \hline
  \hline 
  &  WMAP5 alone & CMB all & CMB all \\
  parameters & ($Y_p$ free) & ($Y_p$ free) & ($Y_p=0.24$)\\
  \hline
  $\omega_b$ & $0.0228\pm0.0006$ & $0.0229\pm0.0005$ & $0.0229^{+0.0006}_{-0.0005}$ \\
  $\omega_c$ &  $0.109^{+0.006}_{-0.009}$& $0.113^{+0.006}_{-0.007}$ & $0.112^{+0.005}_{-0.006}$\\
  $\theta_s$ & $1.040^{+0.004}_{-0.006}$ & $1.043\pm0.004$ & $1.043\pm0.003$\\
  $\tau$ & $0.088^{+0.016}_{-0.018}$ & $0.087^{+0.016}_{-0.018}$ & $0.087^{+0.016}_{-0.018}$\\
  $n_s$ & $0.964^{+0.016}_{-0.018}$ & $0.967^{+0.016}_{-0.015}$  & $0.966^{+0.013}_{-0.014}$\\
  $\ln(10^{10}A_s)$ & $3.06\pm0.06$ & $3.08^{+0.04}_{-0.05}$ & $3.07\pm0.04$\\
  $Y_p$ & $<0.44 (95\%)$ & $0.25^{+0.10}_{-0.07}$ & ---\\
  $A_\mathrm{SZ}$ & $1.1^{+0.9}_{-0.3}$ & $1.1^{+0.9}_{-0.3}$ & $1.0^{+1.0}_{-0.4}$\\
  \hline
  $\Omega_m$ & $0.25\pm0.03$ & $0.27\pm0.03$ & $0.26^{+0.02}_{-0.03}$\\
  $H_0$ &  $72.3^{+2.6}_{-2.8}$ & $71.7^{+2.2}_{-2.6}$ & $71.7^{+2.3}_{-2.5}$\\
  \hline
  \hline 
\end{tabular}
\caption{Mean values and 68\% errors from current observations of CMB
  for the cases with WMAP5 alone and all data combined. (Regarding $Y_p$, 
  an upper bound at 95\% C.L. is given for the case with WMAP5 alone.)
  In the last column, the value of $Y_p$ is fixed as $Y_p=0.24$.}
  \label{table:current}
\end{center}
\end{table}

Next we discuss the effects of the prior of $Y_p$ on the determination
of other cosmological parameters. For this purpose, we repeated a MCMC
analysis fixing the value of the helium abundance to $Y_p = 0.24$ as
in usual analyses. We use all the CMB data (i.e., WMAP5, ACBAR, BOOMERANG
and CBI) here. In Table~\ref{table:current}, in the last column, the
constraints on cosmological parameters for the case with fixing
$Y_p=0.24$ are shown.  When we compare the constraints for the cases
with and without fixing $Y_p$, the central values as well as the
errors at 68\% C.L.  are almost unchanged.  Thus we can conclude that
the usual practice of fixing of $Y_p =0.24$ scarcely affects the
constraints on other cosmological parameters with the current
precision of CMB data.  However, since we can expect more precise
measurements of CMB in the near future, the prior on $Y_p$ may become
important and can affect the constraints on other cosmological
parameters.  We study this issue in the next section.

%%%%%%%%%%%%%%%%%%%%%%%%%%%%%%
\section{Future constraint from Planck}\label{sec:Planck}
%%%%%%%%%%%%%%%%%%%%%%%%%%%%%%

In this section, we forecast a constraint for the Planck experiment
\cite{:2006uk} focusing on the constraint on $Y_p$ itself and how the
prior on $Y_p$ affects the constraints on other cosmological
parameters.  In fact, constraints from the future Planck experiment
from this viewpoint have already been discussed in
Refs.~\cite{Trotta:2003xg,Ichikawa:2006dt} by using the Fisher matrix
analysis.  As mentioned in the introduction, when the likelihood of
cosmological parameters can be approximated by a multivariate Gaussian
function, the Fisher matrix analysis can give a reliable
prediction. However, in practice, the likelihood function deviates
from the Gaussian form.  Furthermore, since the Fisher matrix analysis
can only predict the uncertainty for a fiducial value, it cannot
extract a bias effect (i.e. the estimated central value deviates from
the fiducial value) which is caused by assuming priors on parameters
and possible correlations among parameters.  Thus it may be better to
make a more reliable prediction by using a MCMC method.  
For this purpose, we follow the approach of 
Ref.~\cite{Perotto:2006rj}.

Here we briefly explain the
method of Ref.~\cite{Perotto:2006rj}.
 Observed anisotropies can be
expanded in spherical harmonics and their power spectra of the
coefficients $a_{lm}^{P}$ are composed of signal parts
$C_l^{PP^\prime}$ and noise parts $N_l^{PP^\prime}$:
\begin{equation}
\langle a_{lm}^{P\ *}a_{l^\prime m^\prime}^{P^\prime}\rangle = (C_l^{PP^\prime}+N_l^{PP^\prime})
\delta_{ll^\prime}\delta_{mm^\prime},
\end{equation}
where ${PP^\prime}$ represents three pairs of maps, $TT$, $EE$ and
$TE$.  The signal parts are computed from a fiducial cosmology.  We
assume the cosmological parameters of the WMAP5 mean values for a
power-law $\Lambda$CDM model as a fiducial model.  As for the noise
power spectra, we assume a Gaussian beam and a spatially uniform
Gaussian white noise. $N_l^{PP^\prime}$ are given as the combined
effects from these and can be approximated as
\begin{equation}
N_l^{PP^\prime}=\delta_{PP^\prime}(\theta_\mathrm{FWHM}\sigma^P)^2
\exp\left[l(l+1)\frac{\theta_\mathrm{FWHM}^2}{8\ln2}
\right],
\end{equation}
where $\theta_{\rm FWHM}$ is the full width at half maximum of the
Gaussian beam and $\sigma^P$ is the root mean square of the
instrumental noise.  For the expected data from the Planck experiment, we use three
frequency channels at 100, 143 and 217\,GHz.  We adopt the following
values for the instrumental parameters \cite{Perotto:2006rj}:
$(\theta_{\rm FWHW} {\rm [arcmin]}, \sigma_T [\mu {\rm K}], \sigma_P
[\mu {\rm K}] ) =(9.5, 6.8, 10.9), (7.1, 6.0, 11.4)$ and $(5.0, 13.1,
26.7)$ for $\nu = 100$, 143 and 217\,GHz, respectively.  We assume
other frequency channels are used to remove foregrounds and they are
ideally removed.

Since the anisotropies from both signal and noise are Gaussian
distributed, the likelihood function of the data ${\mathbf a} = \{
a_{lm}^{T, E} \}$ for a theoretical model with parameters
${\mathbf \Theta} = \{ \theta_i \}$ is given by
\begin{equation}
\mathcal{L}({\mathbf a}| {\mathbf \Theta})\propto
\frac{1}{\sqrt{\bar{C}({\mathbf \Theta})}}\exp
\left(-\frac{1}{2}{\mathbf a}^*[\bar{C}({\mathbf \Theta})]^{-1}{\mathbf a}
\right),
\end{equation}
where $\bar{C}(\mathbf{\Theta})$ is a covariance matrix of the theoretical
data. Denoting a covariance matrix of mock data as $\hat{C}$, the
effective $\chi^2$ is given as
\begin{equation}
\chi^2_{\rm eff}=\sum_l(2l+1)f_{\rm sky}\left[
\ln\frac{|\bar{C}_l|}{|\hat{C}_l|}+\hat{C}_l\bar{C}_l^{-1}-2
\right].
\end{equation}
We take $f_{\rm sky}=0.65$ as the expected sky-coverage for the
Planck experiment. The factor $(2l+1)f_{\rm sky}$ represents the effective number
of independent moments obtained from the observation.  For the MCMC
analysis, we include the data up to $l = 2500$.

Before presenting our results, here we comment on possible
contributions from the thermal and kinetic SZ effect.  
We assume that the thermal SZ effect can be precisely estimated from 
the other lower frequency channels of the Planck survey than those used in our analysis, and
can be removed ideally.
The contribution from the kinetic SZ effect on CMB
anisotropy depends on the details of the reionization process.
For somewhat conventional scenario, as argued in \cite{Zhang:2003nr}, 
it would be only about a few percent in the range of
multipoles we make use of, $l\le 2500$, and also
sufficiently smaller than the expected instrumental noise for the
Planck survey. 
Ref.~\cite{Santos:2003jb} argued that ``patchy" reionization would make
it significantly larger but they found that the shape of the power spectrum 
due to the kinetic SZ effect does not depend much on the reionization model.
Then, they concluded that its effect on the determination of the cosmological parameters can be
neglected by marginalizing over the amplitude of the kinetic SZ power spectrum.
We thus neglect the kinetic SZ effect here.

\begin{figure}[htb]
\begin{center}
\scalebox{0.8}{\includegraphics{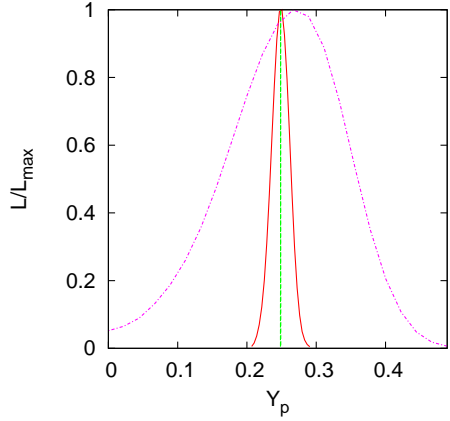}}
\caption{One-dimensional marginalized distributions of $Y_p$.  Shown
  are the distributions from the Planck experiment for the cases with no priors on
  $Y_p$ (red solid line) and assuming the BBN relation (green dashed
  line). For reference, the distribution for the case with no priors
  on $Y_p$ using current CMB data is also shown (dash-dotted magenta
  line).  }
\label{fig:Yp_planck}
\end{center}
\end{figure}

Now we discuss a future constraint on $Y_p$ from the Planck experiment. In
Fig.~\ref{fig:Yp_planck}, one-dimensional marginalized likelihood for
$Y_p$ is shown. For comparison, we also plot the constraint from
current observations.  We can expect that the uncertainty for $Y_p$ at
68\% C.L.  becomes as $\Delta Y_p\sim 10^{-2}$, which is 10 times
smaller than that from current data (see Table~\ref{table:planck}
below).  Since Planck can measure the CMB power spectrum at higher
multipoles very precisely, the effects of damping due to $Y_p$ can be
well probed. It should also be noted that since likelihood functions for $Y_p$ and 
other cosmological parameters have almost the Gaussian form, 
our results here using MCMC approach
are almost the same as those obtained by Fisher matrix analysis 
which has already been done in Refs.~\cite{Trotta:2003xg,Ichikawa:2006dt}. 
Thus we found that the Fisher matrix analysis can give 
a good estimate for Planck data for the parameter set we assumed here.
The results here are consistent with those given in
Ref.~\cite{Perotto:2006rj,Hamann:2007sb} in which a forecast on $Y_p$ is
investigated using MCMC analysis too.

Next we discuss the effects of prior on $Y_p$ on the constraints on
other cosmological parameters in the Planck experiment.  As mentioned
in the Introduction, when one tries to constrain some cosmological
parameters from CMB, the value of $Y_p$ is fixed to be $0.24$ in most
of analysis.  In the previous section, we showed that, when we use
current cosmological data, the fixing of $Y_p=0.24$ does not affect
much the constraints on other cosmological parameters since the value
of $Y_p$ itself is not constrained well.  However, as just shown
above, Planck can measure the value of $Y_p$ much more precisely, thus
we should study the effects of the assumption of $Y_p$ when we
constrain other parameters.  For this purpose, we made MCMC analyses
for three cases: (i) $Y_p$ is not fixed but varied freely, (ii) $Y_p$
is fixed as $Y_p = 0.24$ and (iii) $Y_p$ is regarded as a function of
$\omega_b$ via the standard BBN calculation. For the case (iii), we
relate the value of $Y_p$ to $\omega_b$ by the fitting formula given
in \cite{Burles:2000zk}, to which we refer as ``BBN relation" in the
following. 

\begin{figure}[htb]
\begin{center}
\scalebox{0.6}{
\includegraphics{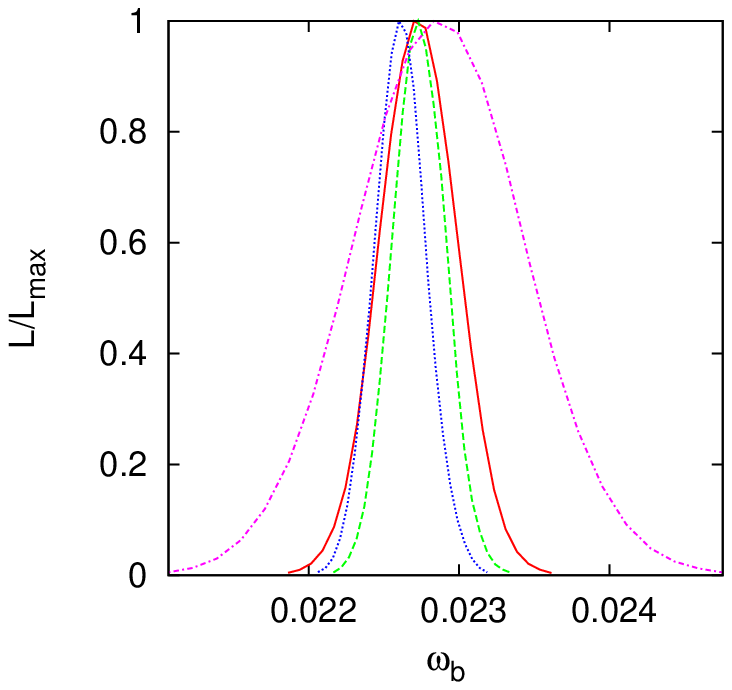}\hspace{-1.2cm}
\includegraphics{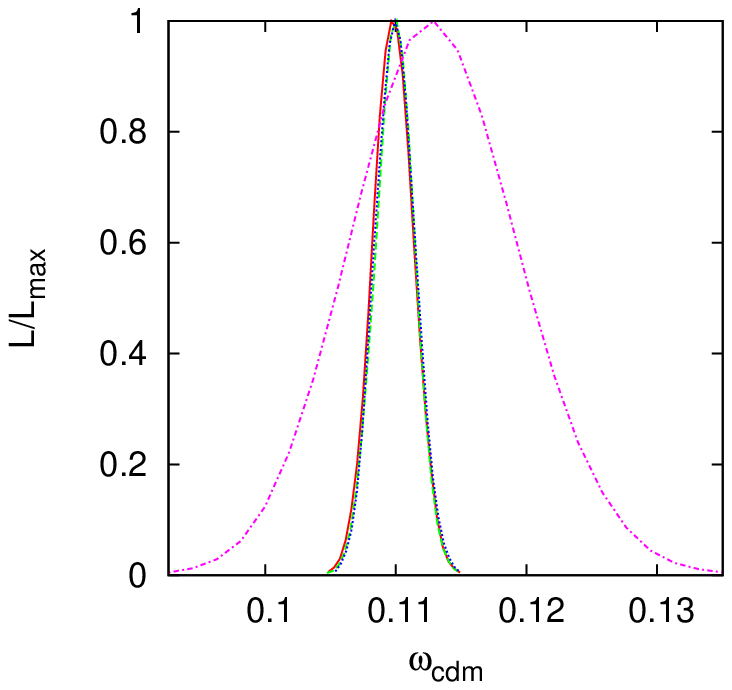}\hspace{-1.2cm}
\includegraphics{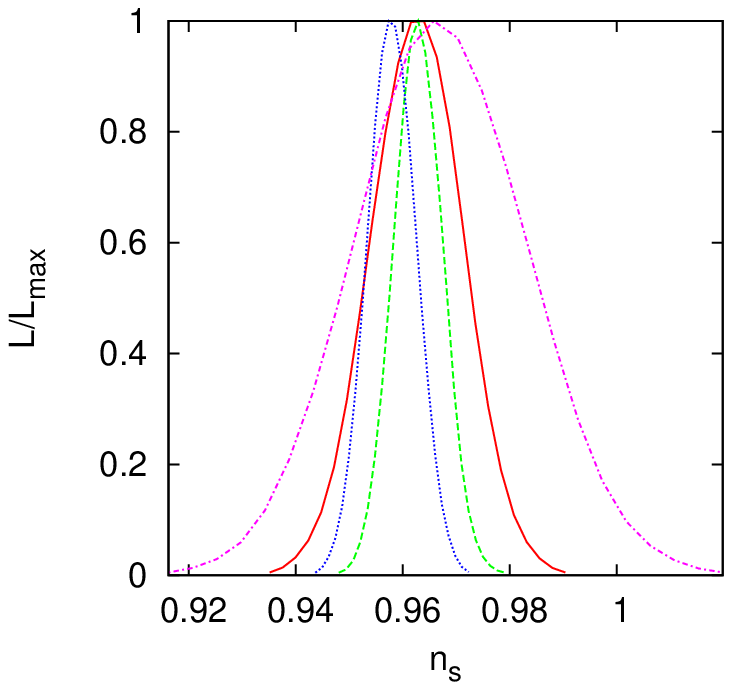}
}
\scalebox{0.6}{
\includegraphics{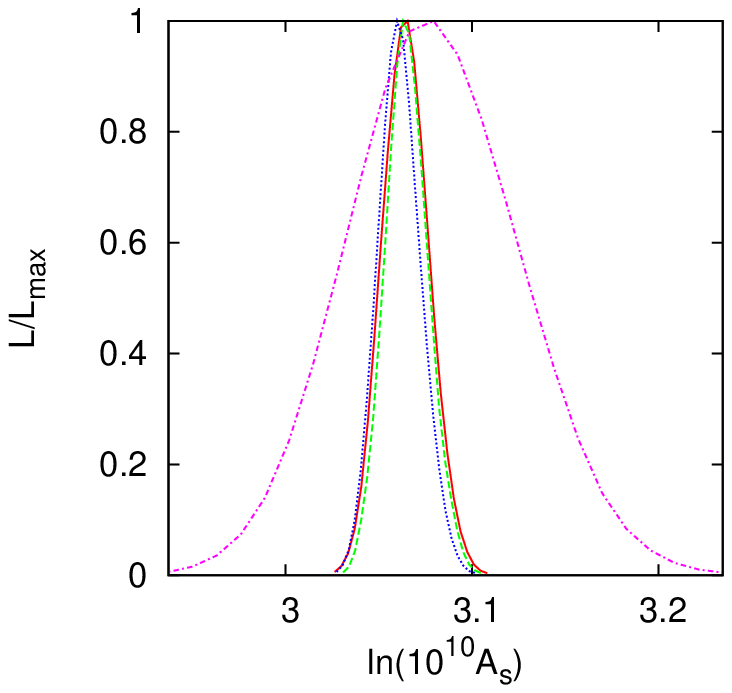}\hspace{-1.2cm}
\includegraphics{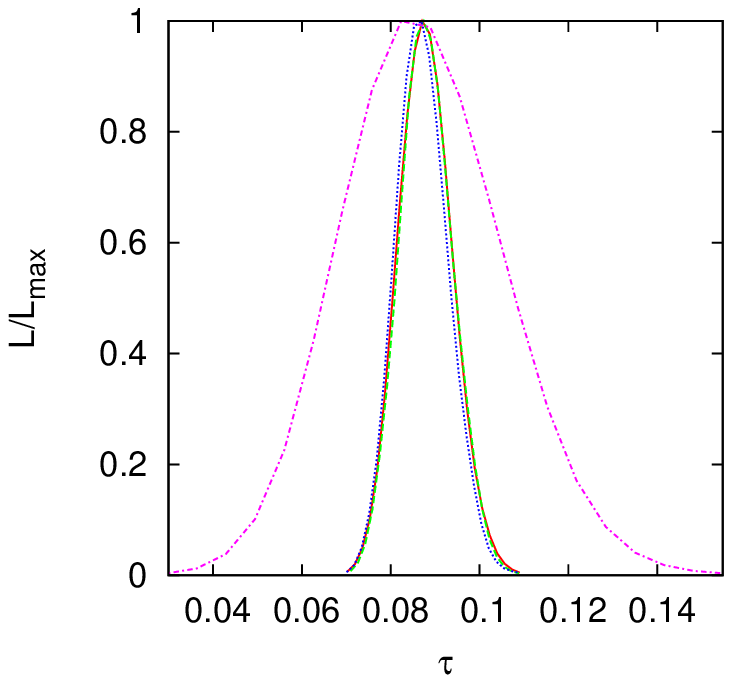}\hspace{-1.2cm}
\includegraphics{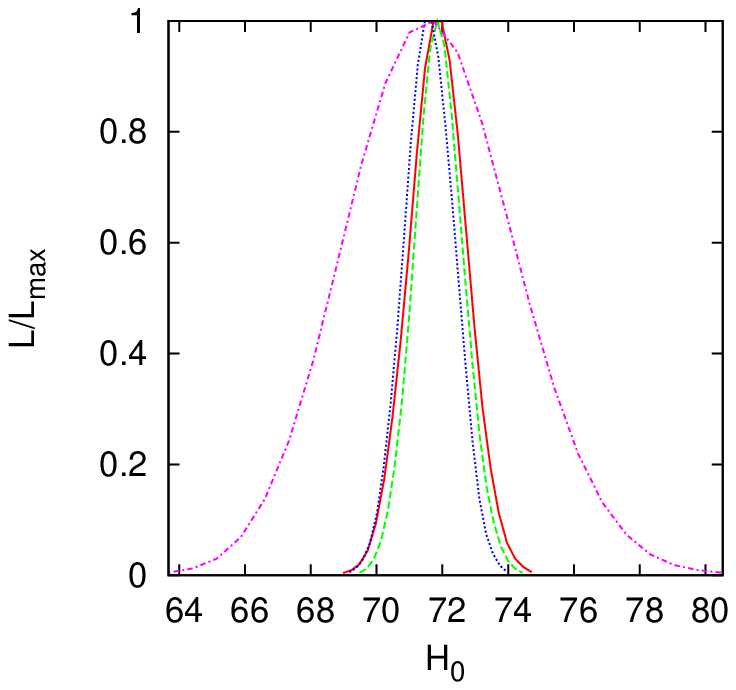}
}
\caption{One-dimensional marginalized distributions of $\omega_b,
  \omega_c, n_s, A_s, \tau, H_0$, using same data as in
  Fig.~\ref{fig:Yp_planck}.  Additionally, the distributions from
  Planck for the case with fixing $Y_p=0.24$ are also shown (dotted
  blue line).}
\label{fig:1Dplanck}
\end{center}
\end{figure}

\begin{table}[ht]
  \begin{center}
  \begin{tabular}{l||r|r|r}
  \hline
  \hline
  parameters & $Y_p$ free & SBBN $Y_p(\omega_b)$ & $Y_p=0.24$\\
  \hline
  $\omega_b$ & $0.02273^{+0.00024}_{-0.00025}$ & $0.02273^{+0.00017}_{-0.00017}$ & 
  $0.02261^{+0.00016}_{-0.00017}$\\
  $\omega_c$ & $0.1098^{+0.0015}_{-0.0014}$ &$0.1099^{+0.0015}_{-0.0014}$ & 
  $0.1100^{+0.0013}_{-0.0016}$\\
  $\theta_s$ & $1.04063^{+0.00057}_{-0.00061}$ & $1.04061^{+0.00037}_{-0.00036}$ & 
  $1.04031^{+0.00034}_{-0.00038}$\\
  $\tau$ & $0.0879^{+0.0054}_{-0.0062}$ & $0.0880^{+0.0055}_{-0.0060}$ & 
  $0.0871^{+0.0049}_{-0.0061}$\\
  $n_s$ & $0.9627^{+0.0079}_{-0.0085}$ & $0.9631^{+0.0046}_{-0.0042}$ & 
  $0.9580^{+0.0042}_{-0.0044}$\\
  $\ln(10^{10}A_s)$ & $3.064^{+0.011}_{-0.013}$ & $3.065^{+0.010}_{-0.013}$ & 
  $3.061^{+0.010}_{-0.012}$\\
  $Y_p$ & $0.248^{+0.014}_{-0.011}$ & $ 0.248586^{+0.000078}_{-0.000076}$ & ---\\
  \hline
  $\Omega_m$ & $0.2567^{+0.0080}_{-0.0086}$ & $0.2565^{+0.0073}_{-0.0080}$ & 
  $0.2587^{+0.0074}_{-0.0083}$\\
  $H_0$ & $71.88^{+0.80}_{-0.85}$ & $71.92^{+0.78}_{-0.66}$ & 
  $71.61^{+0.73}_{-0.72}$\\
  \hline
  \hline
\end{tabular}
  \caption{Mean values and 68\% errors from Planck for some
  assumptions on $Y_p$. }
  \label{table:planck}
\end{center}
\end{table}

Now we show one-dimensional marginalized likelihood for $\omega_b,
\omega_c, n_s, A_s, \tau, H_0$ in Fig.~\ref{fig:1Dplanck}.  In the
figure, three cases (i), (ii) and (iii) are depicted.  In table
\ref{table:planck}, the mean values and errors at 68\% C.L. are shown
for representative parameters.  By looking at Fig.~\ref{fig:1Dplanck},
some features can be noticed.  For $\omega_c$, $A_s$ and $\tau$, the
effects of the prior on $Y_p$ are very small even with the precision
of Planck.  However, for $\omega_b, n_s$ and $H_0$, marginalized
distributions are changed depending on the prior on $Y_p$.  This
tendency can also be seen by reading the errors at 68\% C.L.  from
Table~\ref{table:planck}. For $\omega_b, n_s$ and $H_0$, when we assume the BBN
relation or fix the value of the helium abundance as $Y_p=0.24$, the
errors are reduced to some extent, which clearly indicates that the
assumption of $Y_p$ can affect the determination of other cosmological
parameters.  Furthermore, for these parameters, when we fix
$Y_p=0.24$, the central values differ from the fiducial values by
about the uncertainties at 68\% C.L.\footnote{  
A similar analysis has been done in Ref.~\cite{Hamann:2007sb} recently 
and their results are consistent with ours.}
Therefore, in the Planck era, we
advocate varying the value of $Y_p$ freely in the cosmological
parameter estimation for a conservative constraint, or, if we would
like to do the cosmological parameter estimation in the framework of 
the standard
cosmology, we should impose the BBN relation.

\begin{figure}[htb]
\begin{center}
\scalebox{0.6}{
\includegraphics{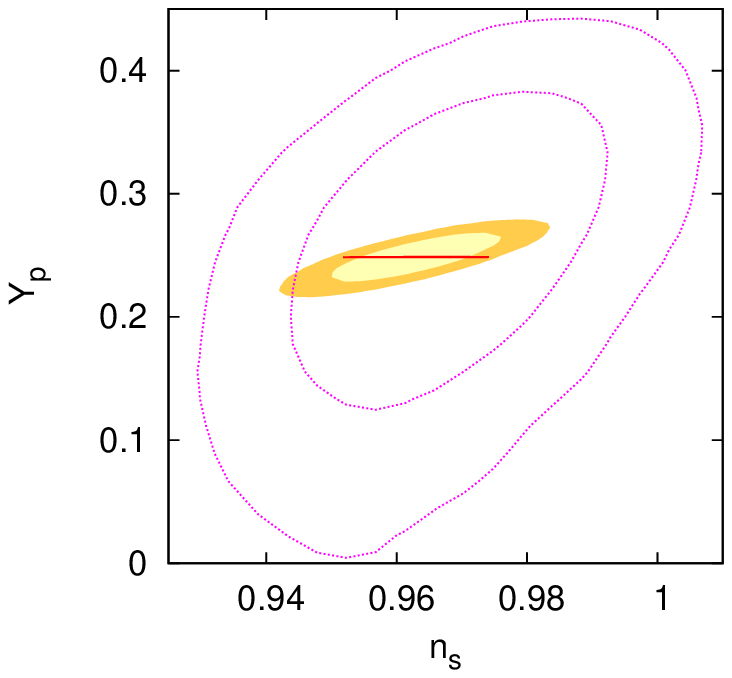} \hspace{-1.7cm}
\includegraphics{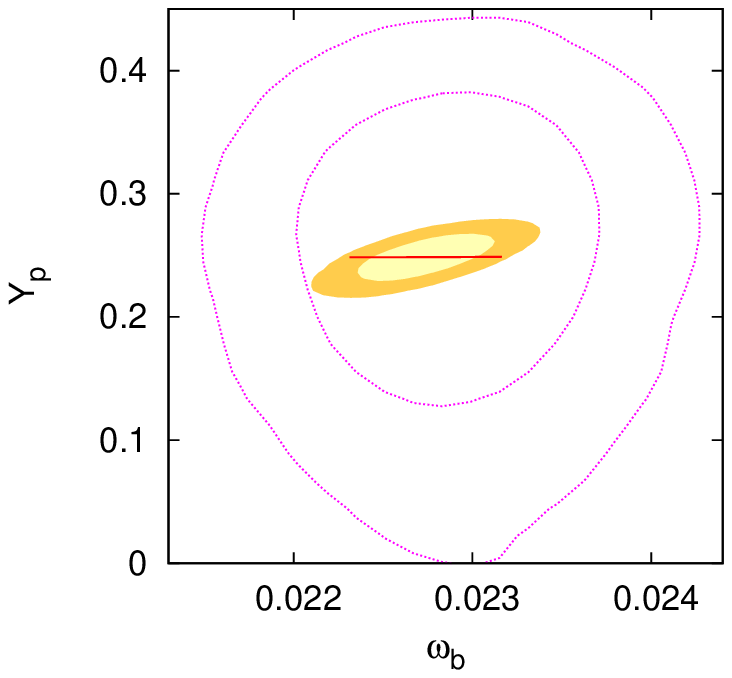}\hspace{-1.4cm}
\includegraphics{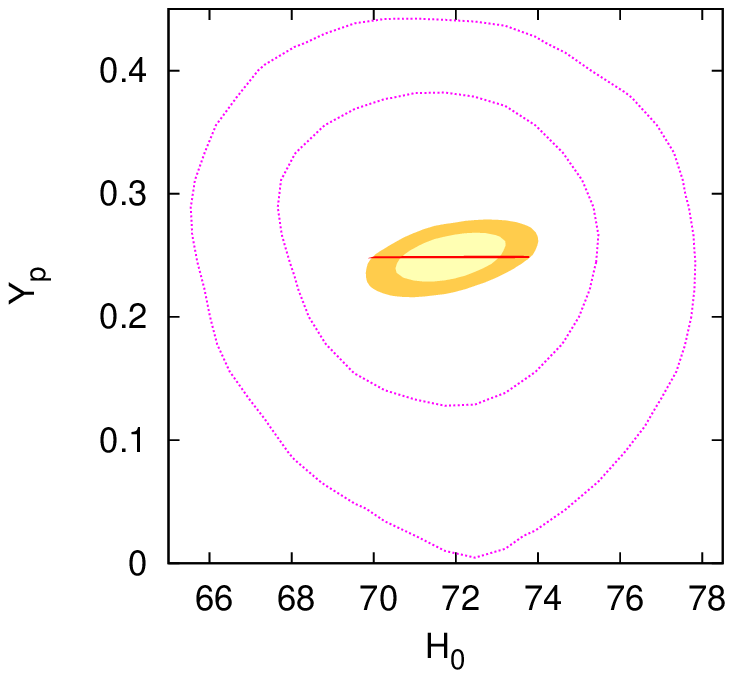}
}
\scalebox{0.6}{
\includegraphics{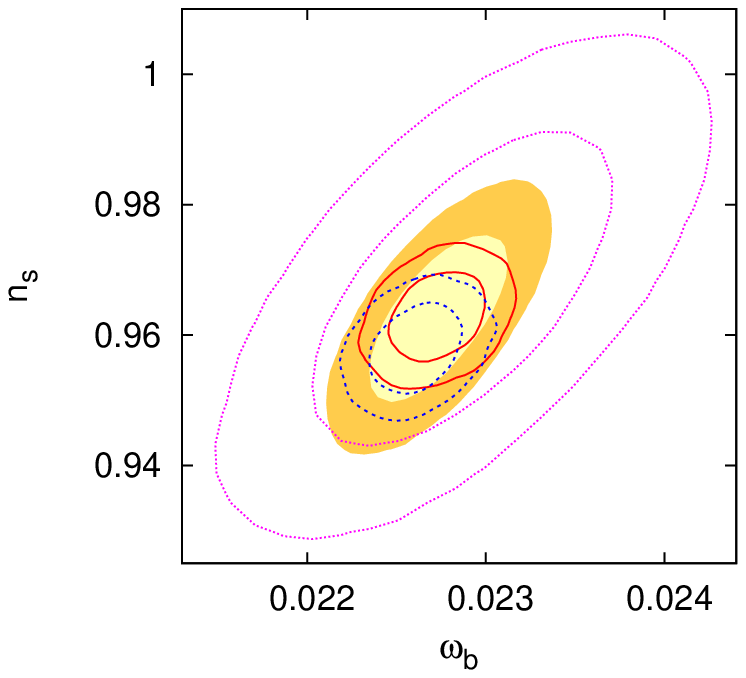}\hspace{-1.5cm}
\includegraphics{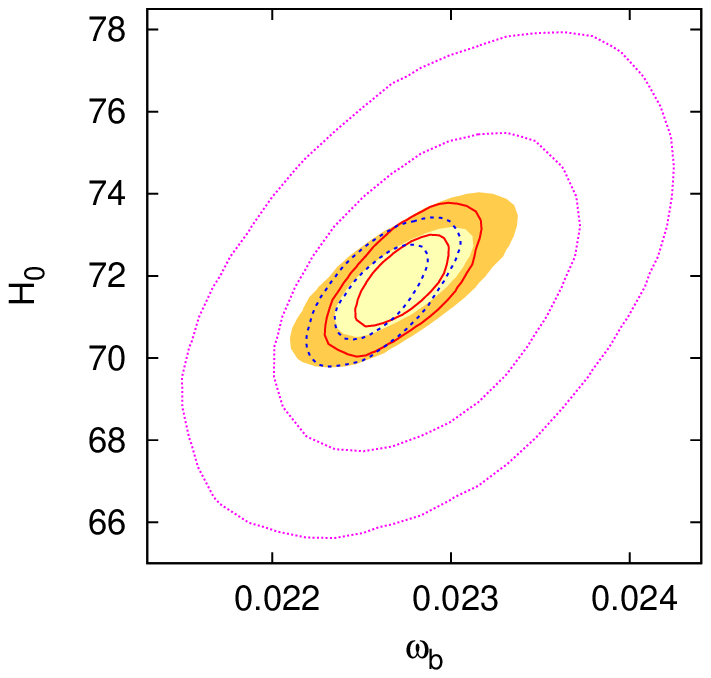} \hspace{-1.5cm}
\includegraphics{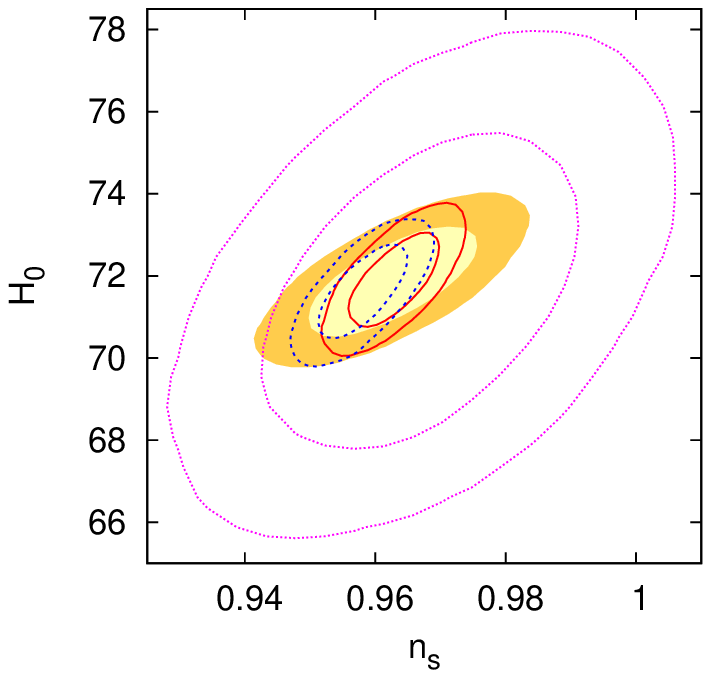}
}
\caption{ Two-dimensional marginalized constraints in several
  combinations of cosmological parameters.  Shown are a Planck
  forecast on the constraints for the cases with no prior on $Y_p$
  (orange and yellow shaded region), assuming the BBN relation (solid
  red line) and fixing $Y_p=0.24$ (blue dotted line).  For reference,
  the constraint for the case with no prior on $Y_p$ using current CMB
  data is also shown (dash-dotted magenta line).  }
\label{fig:2Dplanck}
\end{center}
\end{figure}

To see how these parameters are correlated with $Y_p$, 2D marginalized
contours may be useful, which are shown in
Fig.~\ref{fig:2Dplanck}. From this figure, we can see that $Y_p$ and
these parameters $\omega_b, n_s$ and $H_0$ are positively
correlated. Positive correlations of $Y_p$ with $n_s$ mainly come from
the degeneracy at higher multipoles where the effect of the diffusion
damping is significant. From Eqs.~\eqref{eq:H2}, \eqref{eq:H3},
\eqref{eq:H4} and \eqref{eq:H5}, the response of $H_i$ for $i \ge 2$
to the change of $Y_p$ and $n_s$ are opposite sign, which indicates
that the correlation between these are positive.  Correspondingly,
$\omega_b$ and $Y_p$ becomes positively correlated because of the
positive correlation between $\omega_b$ and $n_s$ which can be read
off, in particular from Eq.~\eqref{eq:H1}.  For the correlation of
$Y_p$ with $H_0$, it should be noticed that the position of the first
peak can be significantly affected by changing $n_s$ and $H_0$. By
increasing the value of $Y_p$, the diffusion damping suppresses the
power on small scales. To compensate this effect to fit the data well,
increasing $n_s$ can enhance the power. Due to the change of $n_s$,
the first peak position is in turn also shifted. However, as is
already well determined by WMAP, the position of the first peak should
be tuned to the right position to fit the data well. This can be done
by changing $H_0$ as can be seen from Eq.~\eqref{eq:L1}.  In fact, the
change of $Y_p$ itself can also shift the peak position. However the
direct effect of $Y_p$ on $l_1$ is very small compared to other
quantities.

\begin{figure}[htb]
  \scalebox{0.5}{
  \includegraphics{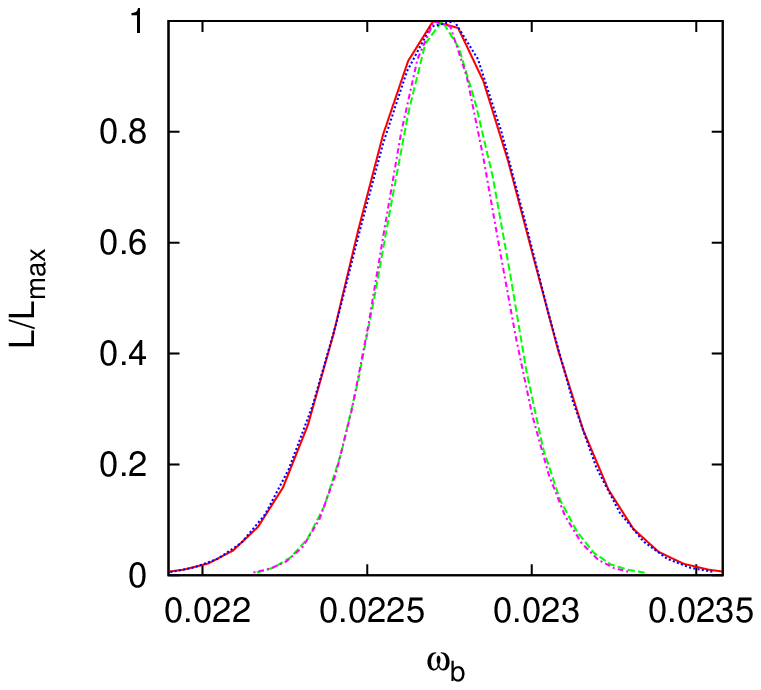} \hspace{-2.9cm}
  \includegraphics{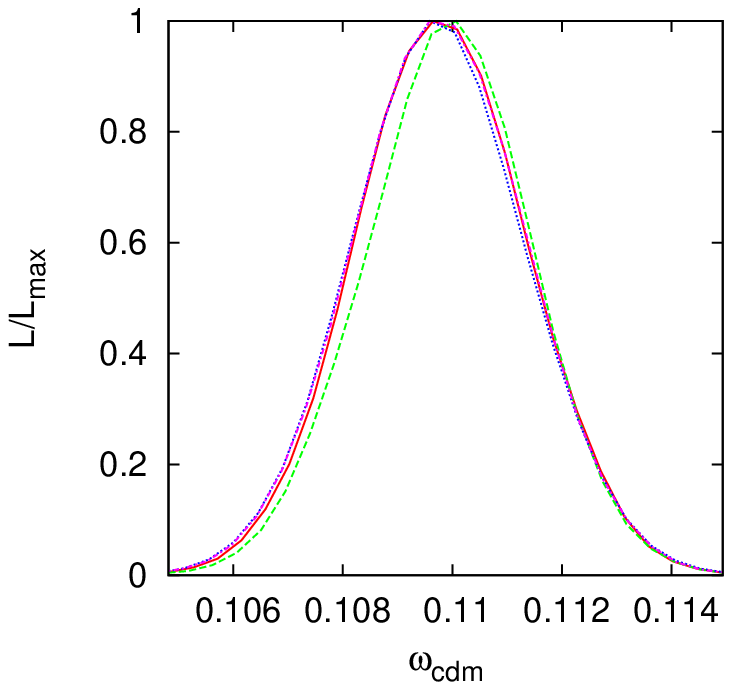}\hspace{-2.6cm}
  \includegraphics{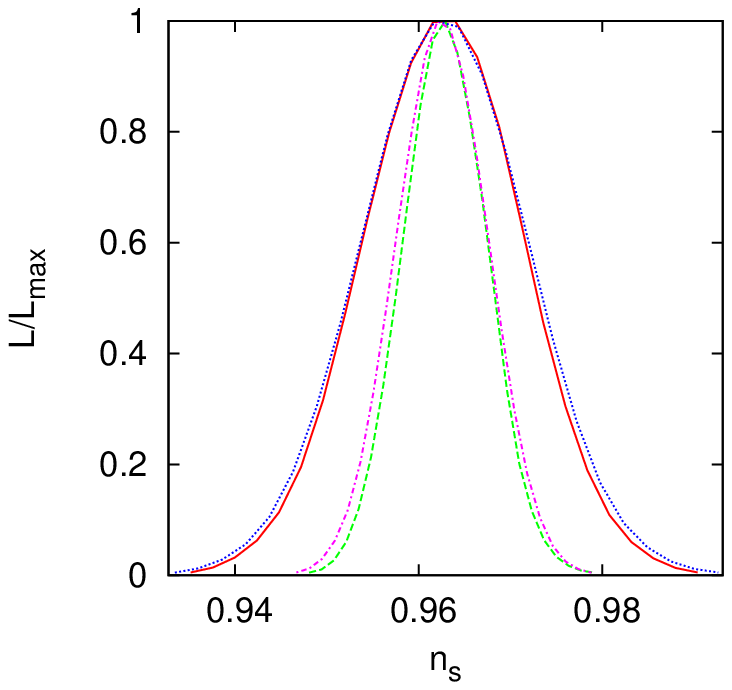}\hspace{-2.6cm}
  \includegraphics{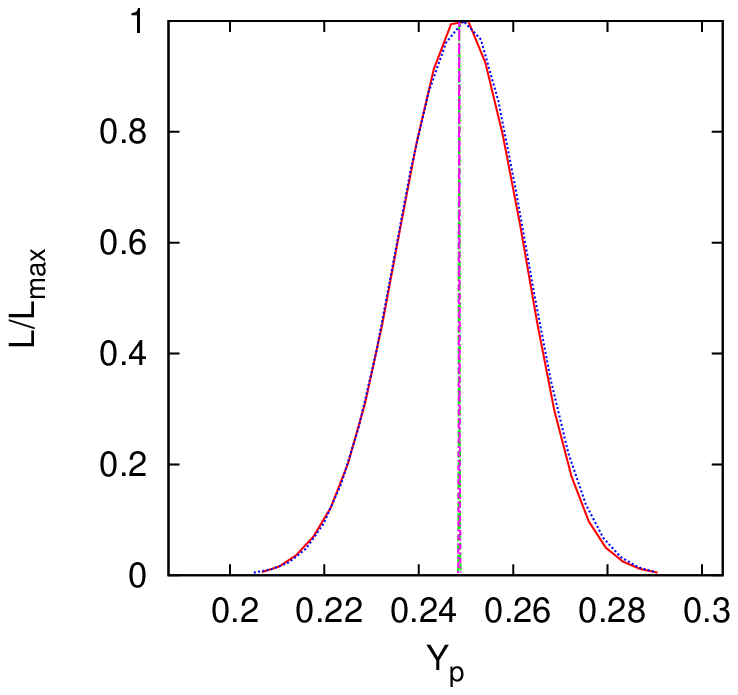}
  }
  \scalebox{0.5}{
  \includegraphics{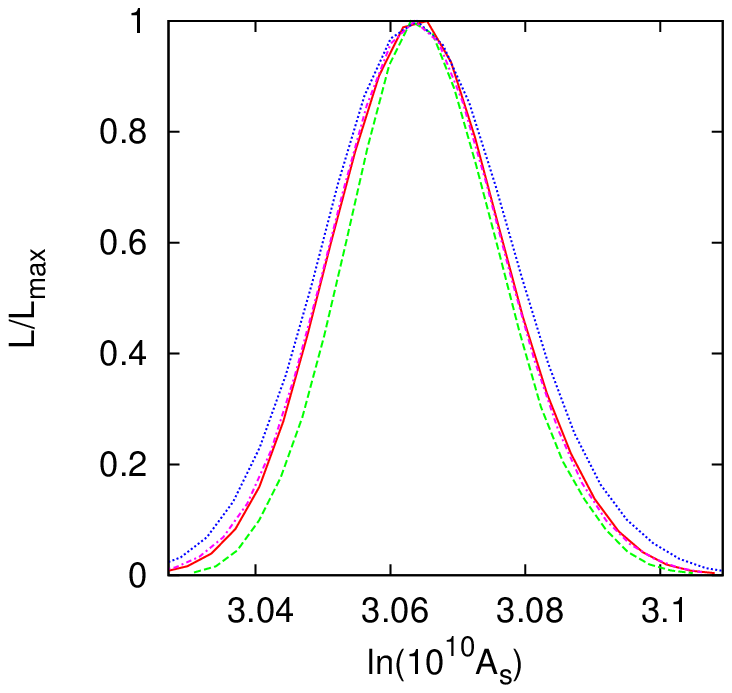}\hspace{-2.9cm}
  \includegraphics{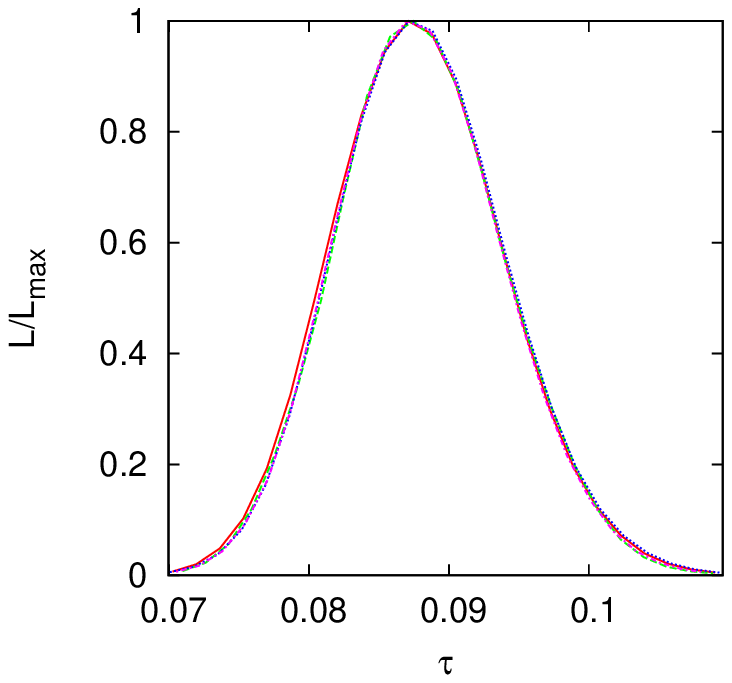} \hspace{-2.6cm}
  \includegraphics{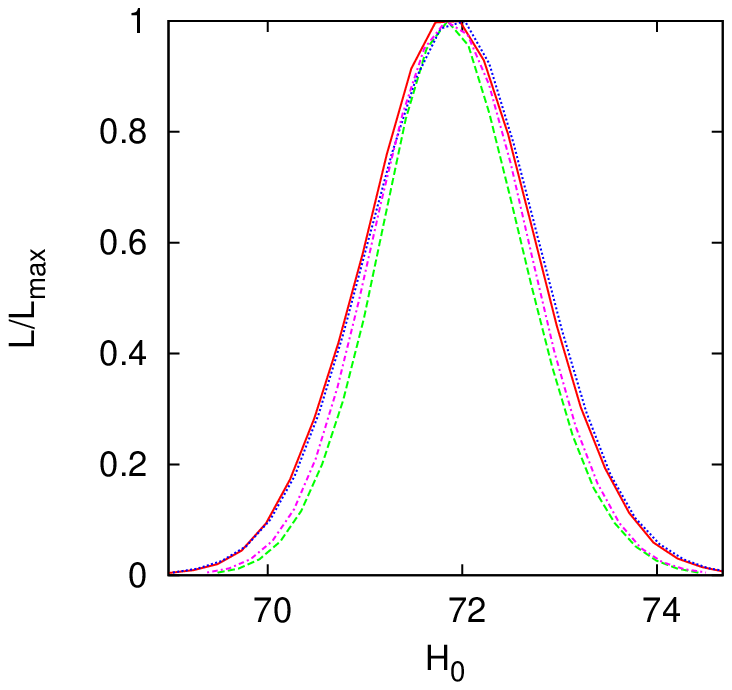}
  }
\begin{center}
  \caption{ Comparison for the cases with and without uncertainties
    from the recombination process being considered. Shown are the
    cases with $Y_p$ being treated as a free parameters and the fudge
    factors being fixed to the standard values (red solid line), $Y_p$
    being treated as a free parameters and the fudge factors being
    treated as free parameters (blue dotted line), $Y_p$ from the SBBN
    relation and the fudge factors being fixed to the standard values
    (green dashed line) and $Y_p$ from the SBBN relation and the fudge
    factors being treated as free parameters (magenta dot-dashed 
    line).  }
\label{fig:fudge}
\end{center}
\end{figure}

Finally, we would briefly discuss how our arguments above are affected
by considering the uncertainties arising from the recombination
process. Several authors have claimed that the uncertainties in the
theory of recombination make some effects on the CMB power spectra and
the determination of the cosmological parameters from CMB
\cite{Lewis:2006ym,Wong:2007ym}.
Although some detailed studies of the recombination modeling have 
been done \cite{Chluba:2005uz,Chluba:2006bc,Chluba:2007yp,Switzer:2007sn,Hirata:2007sp,Switzer:2007sq}, more developments of the recombination modeling 
 may be needed \cite{Wong:2007ym}.
Since the primordial helium abundance, which we are studying in this 
paper, directly affects the recombination history, it may be interesting 
to check how the uncertainties in the recombination modeling affect the 
determination of $Y_p$ and other cosmological parameters in the future 
Planck experiment. 
For this purpose, we treat two numerical
parameters, the so-called fudge factors $F_H$ and $b_{He}$ used in 
{\tt recfast} \cite{Seager:1999bc,Seager:1999km,Wong:2007ym}
as free parameters to represent the uncertainties of the recombination modeling.
$F_H$ is introduced to fit 
the hydrogen recombination rate in the three-level approximation 
to the result from multi-level calculations. 
$b_{He}$ is a fitting parameter for the recombination rate of HeI.
We repeated a MCMC analysis including $F_H$ and $b_{He}$ 
and marginalized over these parameters with the top hat priors given in
Table~\ref{table:priors}, which are very conservative ones.
In Table~\ref{table:fudge}
we summarize the resultant constraints on the cosmological parameters
from expected Planck data. We also show the probability distributions
for several cosmological parameters in Fig.~\ref{fig:fudge}.  
  Table~\ref{table:fudge} shows that, 
 even if we adopt a very conservative prior on the fudge factors, 
  the uncertainties of the recombination 
  modeling which is represented by  $F_H$ and $b_{He}$ 
  do not significantly affect the determination of 
  cosmological parameters in the Planck era. 
  (Errors for some parameters are changed at most by 10\%.) 
Therefore, we can say at least that the theoretical uncertainties of the
recombination process which are discussed recently do not affect 
our previous discussions.
However, since our analysis is limited, we need more understanding of the recombination process
and more detailed analysis in order to reduce systematic errors in the helium estimation from future CMB data.

\begin{table}[ht]
  \begin{center}
  \begin{tabular}{l||r|r|r|r}
  \hline
  \hline
  & \multicolumn{2}{c|}{$Y_p$ free} & \multicolumn{2}{c}{SBBN $Y_p(\omega_b)$} \\
  \hline
  parameters & $F_H$, $b_{He}$ fixed & $F_H$, $b_{He}$ free & $F_H$, $b_{He}$ fixed & $F_H$, $b_{He}$ free \\
  \hline
  $\omega_b$ & $0.02273^{+0.00024}_{-0.00025}$ & $0.2273^{+0.00025}_{-0.00024}$& 
  $0.02273^{+0.00017}_{-0.00017}$ & $0.002273^{+0.00016}_{-0.00018}$ \\
  $\omega_c$ & $0.1098^{+0.0015}_{-0.0014}$ & $0.1098^{+0.0014}_{-0.0016}$ & 
  $0.1099^{+0.0015}_{-0.0014}$ & $0.1098^{+0.0014}_{-0.0016}$ \\
  $\theta_s$ & $1.04063^{+0.00057}_{-0.00061}$ & $1.04066^{+0.00060}_{-0.00061}$ & 
  $1.04061^{+0.00037}_{-0.00036}$ & $1.04064^{+0.00036}_{-0.00039}$ \\
  $\tau$ & $0.0879^{+0.0054}_{-0.0062}$ & $0.0882^{+0.0052}_{-0.0062}$ & 
  $0.0880^{+0.0055}_{-0.0060}$ & $0.0880^{+0.0050}_{-0.0064}$ \\
  $n_s$ & $0.9627^{+0.0079}_{-0.0085}$ & $0.9628^{+0.0091}_{-0.0081}$ & 
  $0.9631^{+0.0046}_{-0.0042}$ & $0.9626^{+0.0046}_{-0.0050}$ \\
  $\ln(10^{10}A_s)$ & $3.064^{+0.011}_{-0.013}$ & $3.064^{+0.013}_{-0.014}$ & 
  $3.065^{+0.010}_{-0.013}$ & $3.064^{+0.012}_{-0.012}$ \\
  $Y_p$ & $0.248^{+0.014}_{-0.011}$ & $0.249^{+0.013}_{-0.012}$ & 
  $ 0.248586^{+0.000078}_{-0.000076}$ & $0.248583^{+0.000087}_{-0.000058}$ \\
  \hline
  $\Omega_m$ & $0.2567^{+0.0080}_{-0.0086}$ & $0.2564^{+0.0081}_{-0.0088}$ & 
  $0.2565^{+0.0073}_{-0.0080}$ & $0.2565^{+0.0080}_{-0.0081}$ \\
  $H_0$ & $71.88^{+0.80}_{-0.85}$ & $71.91^{+0.86}_{-0.82}$ & 
  $71.92^{+0.78}_{-0.66}$ & $71.89^{+0.75}_{-0.75}$ \\
  \hline
  \hline
\end{tabular}
\caption{Comparison for the cases with and without uncertainties from
  the recombination process being considered. In the first two columns
  $Y_p$ is treated as a free parameter and in the latter two columns
  the SBBN relation is assumed.}
  \label{table:fudge}
\end{center}
\end{table}

%%%%%%%%%%%%%%%%%%%%%%%%%%%%%%
\section{Summary}\label{sec:summary}
%%%%%%%%%%%%%%%%%%%%%%%%%%%%%%

We studied the constraint on $Y_p$ and the effects of the priors for
$Y_p$ on constraining other cosmological parameters using current CMB
data from WMAP5, ACBAR, BOOMERANG and CBI, and also from the future
Planck experiment.  After briefly reviewing the effects of $Y_p$ on
CMB, we studied current constraints on the primordial helium
abundance.  We obtained the current limit on $Y_p$ from WMAP5 alone as
$Y_p \le 0.44$ at 95\%\,C.L., which is improved to be 
$Y_p = 0.25^{+0.10 (+ 0.15)}_{-0.07 (-0.17)}$
at 68\% (95\%) C.L.  by adding the data of ACBAR,
BOOMERANG and CBI around the damping tail.  We have also considered
how the prior of $Y_p$ can affects the constraints on other
cosmological parameters using currently available data. We found that,
at the present precision level of CMB measurements, the prior on $Y_p$
has little effect for determinations of other cosmological parameters.

We have also investigated the future constraint from the Planck
experiment.  By performing a MCMC analysis, we derived an expected
error for the helium abundance from the future Planck experiment and
found that it will be well measured with the accuracy of $\Delta
Y_p\sim10^{-2}$ (68\% C.L.)  in the Planck experiment, which is 10 times smaller
values compared with current data. Furthermore, it may be interesting
to notice that this precision is comparable to that obtained by HII
region observations. As for the effects of the prior on $Y_p$ on the
determination of other cosmological parameters, we found that, with
the precision of Planck, the assumption on $Y_p$ can affect the
constraints on other cosmological parameters such as $\omega_b$ and
$n_s$. In this respect, the prior on $Y_p$ can be important for
determining the other parameters. In addition, the constraint on $Y_p$
from CMB itself can be an independent test from other methods such as
using HII region observations.
 
In the near future, we can have more precise measurements of CMB.
Such upcoming data would give us more precise information of the
primordial helium abundance.  At the same time, it is necessary to
study the effects of the helium abundance more rigorously in order to
extract information of other cosmological parameters.

\bigskip
\bigskip

{\sl Note added:} While we were finishing the present work,
Ref.~\cite{Hamann:2007sb} appeared on the arXiv, which has some
overlap with our analysis on the constraints from the future Planck
experiment.

\bigskip
\bigskip

\noindent 
{\bf Acknowledgments:} 
This work is supported in part by the Sumitomo Foundation (T.T.) and
the Grant-in-Aid for Scientific Research from the Ministry of
Education, Science, Sports, and Culture of Japan, No.\,18840010 (K.I.)
and No.\,19740145 (T.T.). T.S. would like to thank the Japan 
Society for the Promotion of Science for financial support.

\end{document}